\documentclass[aps,pre,twocolumn,groupedaddress,showpacs]{revtex4}

\usepackage{graphicx}
\usepackage{color}

\begin{document}


\title{Rotation of a melting ice at the surface of a  pool}


\author{S. Dorbolo, N. Adami, C. Dubois, H. Caps, N. Vandewalle, and B. Darbois-Texier}
\affiliation{
GRASP, D\'epartement de Physique B5, Universit\'e de Li\`ege, B-4000 Li\`ege, Belgium.\\
}


\date{\today}

\begin{abstract}
Large circular ice blocks up to 80 m of diameter have been observed on frozen river around the world. This rare event has been reported in a publication in 1993. This fascinating self-fashioned object slowly rotates at about 1$^o$ per second. In this paper, we report a model experiment consisting in a 85 mm of diameter ice disc at the surface of a thermalised pool. The rotation speed has been found to increase with the bath temperature. Using particle image velocimetry technique, we evidence the presence of a vortex  below the ice block. This vortex results from the descending flow of high density water at 4$^o$C.  The vorticity of the vortex induces the rotation of the ice block.  This mechanism is generic of any vertical flow that generates a vortex which induces the rotation of a floating object. \end{abstract}


\maketitle

\section{Introduction}
Around the world large floating ice blocks have been observed on frozen rivers. This pretty rare phenomenon can be observed during winter news broadcast as the discs surprisingly rotate. In a paper of 1993, Nordell {\it et al} reported several events of rotating ice discs \cite{weather}. The Table I presents the diameters $D$, the rotating speed $w$ expressed in degrees per second for different observations (place, date and references are indicated). The last line concerns the observation along the present work. The size of the discs has been observed to spread over 2 orders of magnitudes from 1 m to 100 m. On the other hand, the rotation speeds do not seem to depend on the size of the disc. One notes that $w$ are of the order of 0.9$\pm$0.4 deg/s.  The mechanism of the disc formation is still to be discovered in detail. Nowadays, two hypothesis are confronted: either the disc is thought to be created by the  aggregation of frazil ice in a vortex generated by the river or an ice block is formed before being rounded \cite{weather}. In the present paper, we study ice discs released at the surface of a pool of water maintained at a given temperature. We discover that the disc starts rotating spontaneously. We found that, while melting, the ice disc generates a vortex under its immersed surface. This vortex is sufficient to make the ice disc rotate.
\begin{table}
\begin{tabular}{l l l l l}
$D$ (m) & $w$ (deg/s) & place & year & ref \\ \hline
2 & 0.57 &   Rancho Nuevo Creek, USA & 1983 & \cite{eos} \\
3 & 1.4 & River Otter, UK & 2009 &Ê\cite{daily} \\
8 & 1.4   & Mianus River, USA & 1895 & \cite{sa}\\
49 & 0.66 & Pite River, Noway & 1987 & \cite{weather} \\
55 & 1.3 & Nidelva River, Norway & 1971 & \cite{hblanche} \\
79.8 & 0.54 & Pite River, Norway & 1994 & \cite{weather} \\
81 & 0.46 & Pite River, Norway & 1994   & \cite{weather} \\
0.085 & 3 (at 20$^o$C) &  this work & &\\
\end{tabular}
\end{table}
\section{Experimental set-up}
The water pool consisted in a large stainless steel circular vessel (300 mm of diameter and 150 mm of depth).  The water pool is plunged in an intermediate bath which temperature is regulated by flowing thermalised water through a copper serpentine (see Fig.1).  The thermalized water comes from a flow thermostat (Julabo F12-ED). Finally, two caps were placed at the surface of the intermediate bath (where a pump mixed the liquid for homogenezation purpose) and over the whole experiment in order to avoid any perturbation by the air environment. 
\begin{figure}[!h]
\centering
\includegraphics[width=8cm]{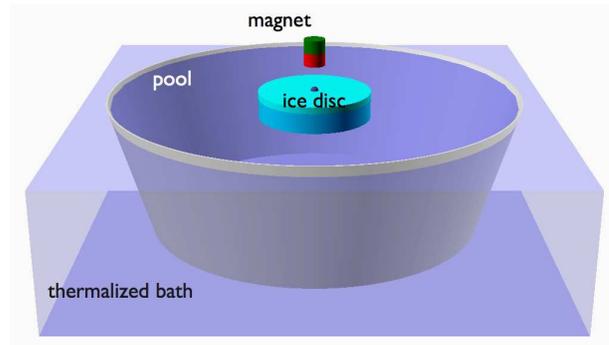}

\caption{Ice disc is placed at the surface of the water pool. The temperature of the water pool is controlled by an intermediated bath of thermalization. The ice disc was either free or magnetically constrained along a vertical axis passing through the center of the disc. In the latter case, a nickel bead was embedded in the ice disc. }
\end{figure}
The ice discs were produced out of circular Petri dish (85 mm of diameter $D$ and 14 mm of height $e$). The Petri dish were embedded in polystyrene so as  the bottom part of the petri dish is not covered. Then the polystyrene frame containing petri dish filled with water are placed in a freezer (-32$^o$C). As the water starts freezing by the bottom of the Petri dish, bubbles and dust are pushed towards the surface of the water. Bubble free ice discs were consequently obtained. 

A dark ellipse (about 50 mm long and 20 mm large) was put on the floating ice disc. By image analysis, the position of the center of the ellipse and its angular position were measured which allows to determine the motion of the ice disc during the melting. The images were captured from a webcam and treated in real-time using a Python program and the OpenCV library. 
\begin{figure}[!h]
\centering
\includegraphics[width=9cm]{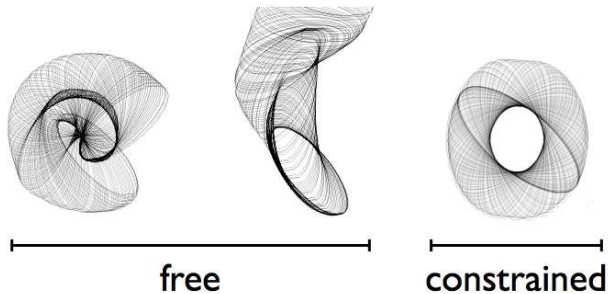}
\caption{(left and centre) Top view trajectories of the motion of the ice disc when the disc is free floating. (right) Trajectory of a ice disc when it is constrained to rotate about its centre. The large axis of the ellipse is 50 mm and the temperature of the thermalised pool 20$^o$C. }

\end{figure}

While melting, one observes that the ice discs move at the surface of the bath. Sometimes they were rotating similarly to the spinning ice discs observed in Rivers. Sometimes they were rotating and translating at the surface of the bath. The ellipse contours have been superimposed to deduce the motion of the discs (1 second separates each successive ellipse). Three typical trajectories are presented in Fig. 2, from left to right: a rotation, a combination of two rotations and the combination of a translation and a rotation.  In order to increase the reproducibility, we suppressed the translational degrees of freedom by constraining the ice disc along a vertical axis crossing the disc at its center. A small nickel bead (2 mm of diameter) was embedded at the bottom center of the ice discs before freezing.  A fixed magnet was placed above the surface of the bath and attracted the nickel bead and consequently the ice disc. Note that the nickel was chosen because (i) the nickel does not oxidise and (ii) the nickel is ferromagnetic with a small remanent field at 273 K.  Indeed, if a magnet was placed in the ice disc instead of the Ni-bead, a torque on the disc may result in the perturbation of the disc motion. 
\begin{figure}[!h]
\centering
\includegraphics[width=8cm]{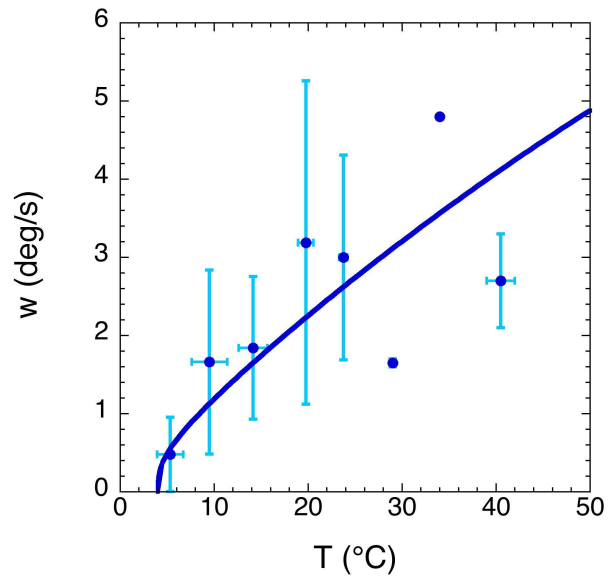}
\caption{Rotation speed of the constrained ice disc expressed in degrees by second as a function of the pool temperature. The point are obtained by averaging the data over bins of 5$¡C$. The solid line is a fit from Eq. (8).}

\end{figure}

\section{Observations and results}
In Fig.3, the rotation speed $w$ [$^o$/s] is plotted as a function of the temperature of the pool. We observe that, even constrained about a vertical axis, the data is rather spread. The data were binned by 5$^o$C boxes and averaged (big bullets in Fig.3). One observes a clear increase of the rotation speed with the temperature of the pool. The rotation  starts to be efficient for a temperature larger than 5$^o$C. This fact indicates that the origin of the rotation is related to the so-called anomaly of the water density behavior with the temperature. Indeed, the density of the water is maximal at a temperature of 4$^o$C.
\begin{figure}[!h]
\centering
\includegraphics[width=8cm]{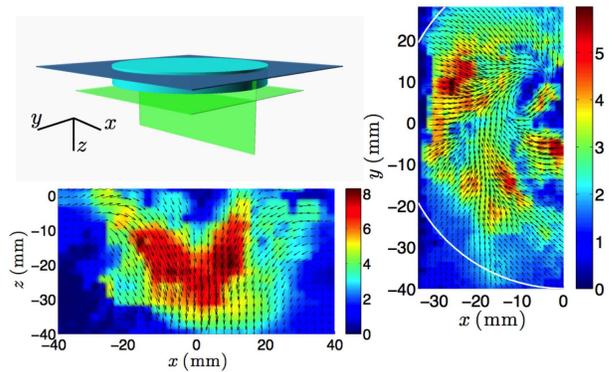}
\caption{Particle Imaging Velocimetry measurements at $T=20^o$C. The sketch presents the orientation of the planes along which the PIV measurements were performed, namely one parallel and one perpendicular to the bath surface (see text). The results for the parallel and the perpendicular planes are presented on the bottom left and on the right of the Figure respectively. The scales indicates the speed in [mm/s].  }

\end{figure}
In order to visualize the motion of the fluid below the ice disc, we colorised the water used to make the discs. This simple experiment allows us to observe the formation of a vortex below the ice disc. The radius of the vortex is typically half that of the disc. 

Quantitative results were obtained by using Particule Image Velocimetry (PIV) technique from DantecDynamics. Fluorescent particles were inserted in the liquid of the bath contained in a transparent tank. For this part, the bath temperature was 20$^o$C. A green LASER of excitation was used and a Phantom camera recorded the motion of the particules. Again, the ice disc was constrained to rotate about one vertical axis defined by using a magnet. The PIV investigations were performed along two perpendicular planes: (i) one plane was taken parallel to the bath surface about 5 mm below the bottom surface of the ice disc and (ii) one plane was take perpendicular to the bath surface intercepting the axis of rotation of the ice disc (see Fig. 4). 

In Fig. 4, the velocity fields are presented for both concerned planes (parallel $x-y$ and perpendicular $x-z$). First, concerning the vertical plane $x-z$, we observe that the fluid moves towards the center of the ice disc located around $x=0$ mm. The maximum speed observed at 1 mm below the bottom of the ice disc ($z=0$ mm) is about 10 mm/s downwards. The analysis of the horizontal plane $x-y$ also indicates maximum speed  of the same order of magnitude. Moreover, a rotation of the flow is observed below the ice disc; in the present case at $x=-10$ mm and $y=$10 mm. Regarding these observations, we conclude that a downwards flow is generated below the ice which creates a vortex. The latter induces the rotation of the disc.  This is supported by the fact that when the level of the water in the pool is low (3 cm), the ice disc does not rotate. 

The proposed mechanism is the following. The ice disc melts and decreases locally the water temperature close to the bottom side of the disc. Eventually, the water reaches locally the temperature of 4$^o$C. At this temperature the density of the water is the highest and the fluid element plunges in the pool. This process is the symmetrical situation to the natural convection. Indeed, this is also supported by the evaluation of the Rayleigh number $$Ra=\frac{g \alpha \rho_0 L^3 \Delta T }{\eta D}$$ where $\alpha$, $D$, $\eta$ and $\rho_0$ are the thermal expansion (88.10$^{-6}$/K), the thermal diffusivity (1.5.10$^{-7}$ m$^2$/s), the kinetic viscosity (10$^{-3}$ Pa s) and the density (1000 kg/m$^3$) of the water respectively, $g$ the gravity acceleration and $L$ ($=R$) the characteristic length of the system. One finds $Ra \approx 8.9\ 10^{6}$. The plume is similar to plume observed in soap films \cite{adam1,adam2}Consequently, regarding the PIV measurement, a model of thermal plume can be applied. In Ref.\cite{book}, the buoyancy flow $F$ is defined as \begin{equation}
F=\frac{\alpha g}{\rho_0 C_p} Q_T
\end{equation} where $\rho_0$ and $C_p$ are respectively the density of water  and the heat capacity at constant pressure and $Q_T$ is the heat flux. Taking into account the heat, the mass and the momentum budget on thin slide of plume, Cushman-Roisin finds an expression for the velocity $v_z$ of the fluid at the center of the plume \begin{equation}
v_z \propto \left (\frac{F}{z}\right )^{1/3}\end{equation} where $z$ the vertical position $z=0$ being the location of the ice disc bottom face.
 \begin{figure}[!h]
\centering
\includegraphics[width=8cm]{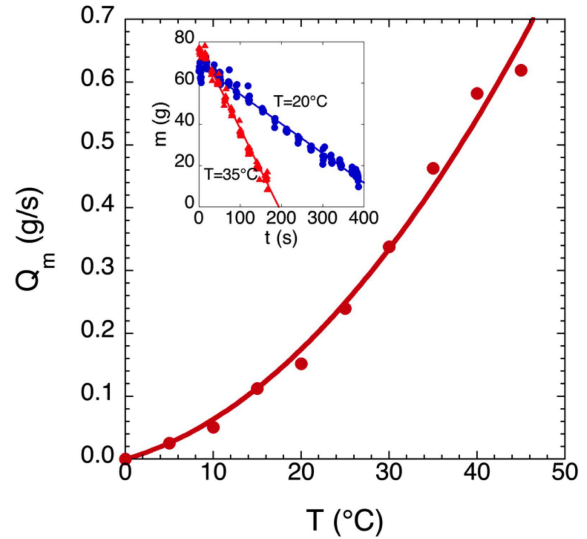}

\caption{Mass flux of the ice disc as a function of the temperature. The inset present the mass decrease of two ice discs as a function of time for two different temperature, i.e. $T=20^o$C (blue bullets) and $T=35^o$C (red triangles). }

\end{figure}
The thermal flux $Q_T$ can be evaluated by determining the flux of mass $\dot m$ of the ice disc if we consider that the heat is used to change the ice into water plus the heat necessary to heat the ice from -30$^o$C to 0$^o$C,\begin{equation}
 Q_T=(\mathcal{L}+C_{ice}\Delta T)\dot m
\end{equation} where $\mathcal{L}$ and $C_{ice}$ are the latent heat of the fusion and the specific heat of ice respectively. The mass of the ice disc as been recorded during the melting. The procedure consisted in weighting the ice disc as a function of time by briefly pulling the disc out of the pool using a string attached to a force sensor.  In the inset of Fig. 5, the mass $m$ of the ice disc is presented as a function of time when the pool is at 20$^o$C (red triangles) and at 35 $^o$C (blue circles). The evolutions of $m(t)$ were fitted by a linear trend $Q_m t$ for different temperatures of the pool. The angular coefficients $Q_m$ are reported as a function of the pool temperature in Fig. 5. In the considered range of temperature, one finds that $Q_m = 3.9\ 10^{-6} T+ 240\ 10^{-9} T^2$. With the assumption that the vertical speed is due to the water anomaly we found that 
\begin{equation}
 v_z\propto \left (\frac{\alpha(T) g}{z \rho_0 C_p}Q_m(\mathcal L +C_p \Delta T)\right ) ^{1/3}
 \end{equation} taking $\alpha=14.10^{-6} (T-4)-97.10^{-9}(T-4)^2$ (in the range 4$^o-50^o$C) \cite{tool}, $\rho_0$=1000 kg/m$^3$, $C_p$=4202 J/kg/K, $C_{ice}$=2060 J/kg/K, $T$=20 $^o$C (the difference of temperature between the ice/water transition and the pool) and $z=2$ cm as in the PIV measurement, one finds $v_z\approx 11 $ mm/s in pretty good agreement with the PIV measurements even the effect of the thermal gradient is not taken into account.

\begin{figure}[!h]
\centering
\includegraphics[width=8cm]{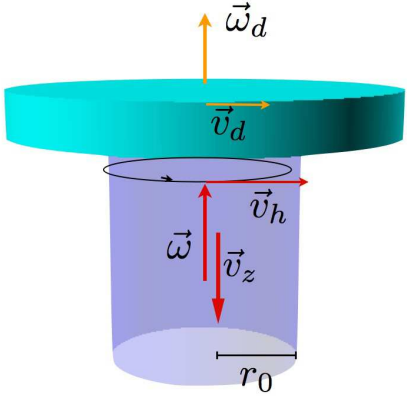}
\caption{Sketch of the side view of the melting ice disc. A vertical plume conducts to a vertical motion of the liquid with a vertical speed $v_z$. The vertical motion of the fluid generates a vortex with an angular moment $\omega$ which entrains the rotation of the ice disc with an angular momentum $\omega_d$.}

\end{figure}
 
 The plume causes the vertical flow which generates a vortex. The relation between the vertical flow and the vortex velocity is rather complex (see for example \cite{lewellen}). In the present case, the vertical flow is driven by natural convection. Our approach is phenomenological as we observed that the vertical speed of the plume is of the same order of magnitude than the vortex. Moreover, we observe that the vortex radius $r_0$ is half the radius $R$ of the ice disc. The vortex angular momentum $\omega$ entrains the rotation of the disc $\omega_d$. The rotation of the disc is postulated to be a viscous entrainment on a boundary layer of a thickness $\delta$.  As a consequence, the driving force $f$ is given by \begin{equation} \label{mot}
 f\simeq\int_0^{r_0}\int_0^{2\pi}\eta \frac{(\omega-\omega_d) r}{\delta}r d\theta dr=2 \pi r_0^3\frac{\eta (\omega-\omega_d) }{3 \delta}
 \end{equation}  with $\eta$ is the viscosity of the water. On the other hand, the friction of the disc with the pool limits the rotation. This friction occurs along the lateral surface of the disc and on the remaining surface on the bottom side of the disc. Taking the same argument as for the entrainment, the friction force is due to viscosity on a layer of thickness $\delta$ reads 
 \begin{equation}
 \begin{array}{  c c c } \label{frein} 
 f_d & =& \int_0^{2\pi} \eta \frac{\omega_d R e}{\delta} d\theta+\int_{r_0}^{R}\int_0^{2\pi}\eta \frac{(\omega-\omega_d) r}{\delta}r d\theta dr \\ f_d & =& 2\pi \eta \omega_d R^3 e/\delta \left (\frac{e}{R}/\frac{1}{3}\right ) \end{array} \end{equation} where $e$ is the thickness of the ice disc. As the rotation speed of the disc is constant, the forces balance and we obtain a relation for the coupling $\gamma$ \begin{equation}
 \gamma=\frac{\omega_d}{\omega}=\frac{1}{1+3\frac{R^2e}{r_0^3}}
 \end{equation} Applied to the present case, one finds that $ \frac{\omega_d}{\omega}\approx  0.06$ with $e=14$ mm.  
 \begin{figure}[!h]
\centering
\includegraphics[width=9cm]{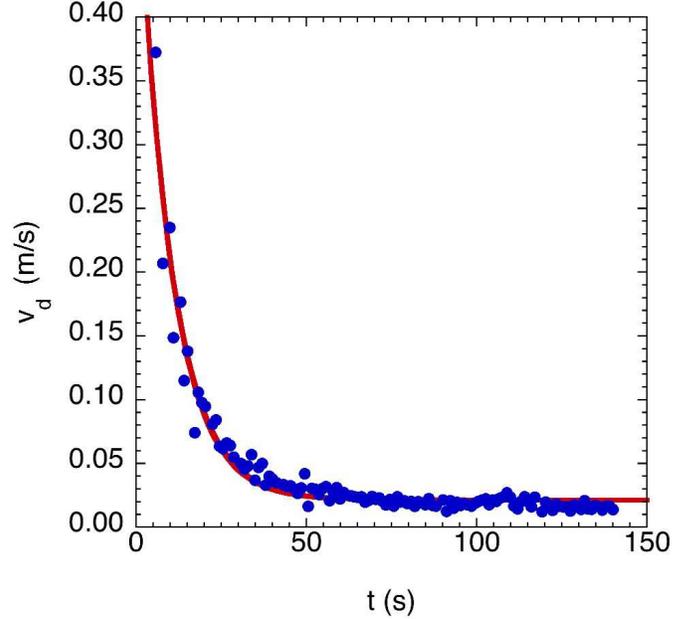}
\caption{Tangential linear speed $v_d$ of the disc has a function of time when the disc was initially spend by hand. The plain line is a fit by a decreasing exponential.}

\end{figure}
 The layer thickness $\delta$ can be evaluated as following. Ice discs were spinned by hand. The rotation motion has been recorded and consequently, the linear speed $v_d=\omega_d R$ were computed as a function of time.  In Fig. 7, the tangential speed of the disc is reported as a function of time. The data were fitted by a decreasing exponential which provides the relaxation rate $\kappa \approx 0.1$ s$^{-1}$.  This behavior indicates that the tangential speed may remodelled by a differential equation $m \dot v_d=-m \kappa v_d$ where $m$ is the mass of the ice disc. The decrease of the speed comes from the friction $f_{hd}$ of the disc (lateral surface and bottom side) with liquid. The calculation is similar to Eq.(6). The integral is performed from $0$ to $R$ instead of between $r0$ and $R$. One obtains 
 \begin{equation}
 f_{hd}=\frac{2 \pi \eta \omega_d R^3}{\delta}\left (e/R+1/3 \right ) = m\kappa \omega_d R \end{equation} One can then determine the value of $\delta\approx 233 \mu$m. 
 
 Finally, we make the assumption that the horizontal speed responsible for the vortex and then for the rotation of the ice disc scales similarly to the vertical speed  $v_z$. In so doing, one finds that
 \begin{equation}
\omega_d =A  \gamma (F/z)^{1/3} \end{equation} The blue curve in Fig. 3 is the fit to the data using this scaling where $A$ is a free parameter and $z=0.01$ m. The agreement is fairly good as only the aspect ratio of the vortex has been left free. Note also that the viscosity of the water does not intervene in the final description of the rotation speed Eq.(8). Indeed, as the propulsion mechanism and the friction mechanism have the same physical origin, the role of the viscosity cancels. 
   
\section{Conclusion}
The spontaneous rotation of an ice disc has been studied at the laboratory scale. The proposed mechanism is that the ice disc cooled the water of the pool. As the water reaches the temperature of 4$^o$C, the water sinks generating a downwards plume and a subsequent downwards flow. This flow trigs a vortex that entrained the ice disc in rotation through a viscous process. This allows to understand why such a process cannot be observed in lakes because the water deep in the lake is at 4$^o$C while the water in between the ice and the bottom is between 0 and 4$^o$C. The thermal plume cannot born. In the rivers, the situation is different as the flow is sufficient to generate the rotation. Finally, one also understands why melting iceberg does not rotate as the melting results in diluting salted sea water. The density inversion cannot be triggered. This mechanism is generic. For example, a caramel disc has been attached to a floating circular boat. The melting of the caramel has generated the rotation of the boat.

\begin{acknowledgments}
SD thanks F.R.S.-FNRS for financial support. This research has been funded by the Interuniversity Attraction Pole Programme (IAP 7/38 MicroMAST) initiated by the Belgian Science Policy Office and by ARC SUPERCOOL.
\end{acknowledgments}


\end{document}